\documentclass{iaus}
\usepackage{graphicx}

\title[Mergers in Early-Type QSO Host Galaxies and Ellipticals]
{Searching for Mergers in Early-Type QSO Host Galaxies and a Control Sample of Inactive Ellipticals}

\author[Bennert et al.]   
{Nicola Bennert$^1$, Gabriela Canalizo$^1$, Bruno Jungwiert$^{1,2}$, Alan Stockton$^3$, Fran\c{c}ois Schweizer$^4$, Chien Peng$^5$, Mark Lacy$^6$}

\affiliation{$^1${Institute of Geophysics and 
Planetary Physics, UC Riverside, CA 92521, USA\\[\affilskip]
$^2$ Astronomical Institute, Academy of Sciences,
Bo{\v c}n\'\i\ II 1401, 141 31 Prague 4, Czech Republic\\[\affilskip]
$^3$ Institute for Astronomy, University of Hawaii, 2680 Woodlawn Dr., 
Honolulu, HI 96822, USA\\[\affilskip]
$^4$ Carnegie Observatories, 
813 Santa Barbara Street, Pasadena, CA 91101, USA\\[\affilskip]
$^5$ Space Telescope Science Institute, 3700 San Martin Drive, 
Baltimore, MD 21218, USA\\[\affilskip]
$^6$ Spitzer Science Center, 
California Institute of Technology, Pasadena, CA 91125, USA}}

\pubyear{2007}
\volume{245} 
\pagerange{100-101}
\date{?? and in revised form ??}
\jname{Formation and Evolution of Galaxy Bulges}
\editors{M. Bureau, E. Athanassoula, \& B. Barbuy, eds.}
\begin{document}

\maketitle

\begin{abstract}
We present very deep HST/ACS images of five QSO host galaxies, 
classified as undisturbed ellipticals in earlier studies.
For four of the five objects, our images reveal strong signs of interaction
such as tidal tails, shells, and other fine structure,
suggesting that a large fraction of QSO host galaxies may have
experienced a relatively recent merger event.
Our preliminary results for a control sample of 
inactive elliptical galaxies do not reveal comparable fine structure.
\keywords{galaxies: active; galaxies: ellipticals; galaxies: interactions; quasars: general; quasars: individual (OX 169, PHL 909, 
PKS 0736+01, PG 0923+201, MC2 1635+119)}
\end{abstract}

\begin{figure}
\includegraphics[width=13.5cm]{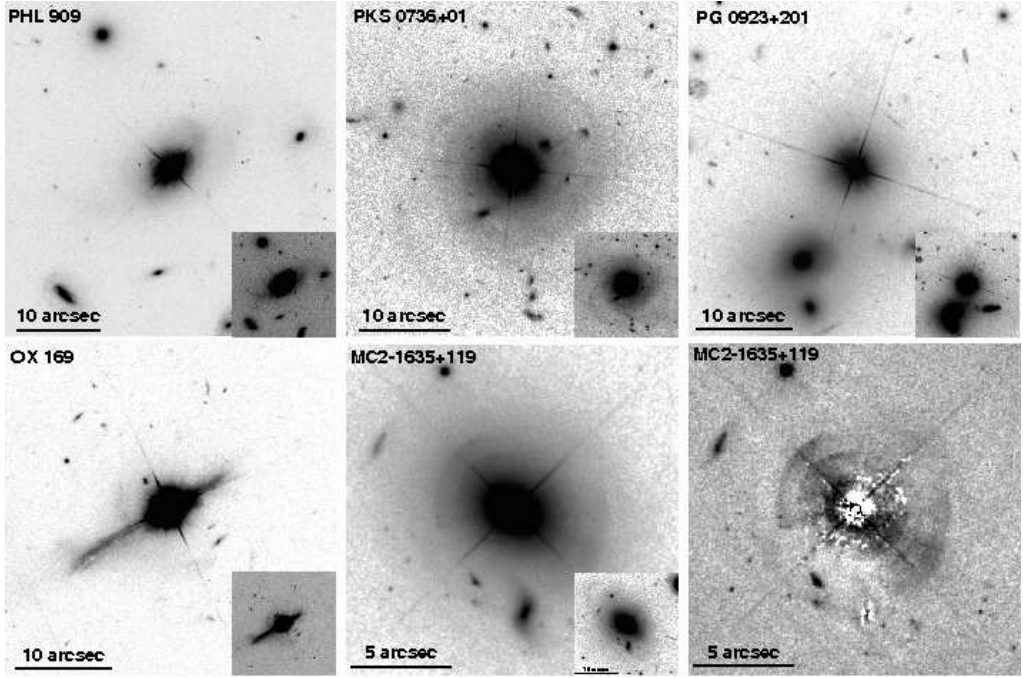}
\caption{Deep HST/ACS images of five early-type QSO host galaxies.
North is up, east is to the left.
{\bf PHL\,909} ({\it top left}): A ring-like structure is seen
around the QSO nucleus. Diffuse outer material form another ring and tidal
tails to both sides of the host.
{\bf PKS\,0736+01} ({\it top middle}): A large ($r$ $\sim$ 50\,kpc) but faint spiral-like
structure surrounds the QSO. The changes in pitch angle may indicate spatial wrapping
of material from a merger event rather than a spiral disk seen face on.
{\bf PG\,0923+201} ({\it top right}): No fine structure can be seen in the host galaxy,
but it lies in an environment with several interacting companions. {\bf OX\,169} ({\it bottom left}):
The extended linear structure is likely a tidal tail seen nearly edge on. Note the extended shell-like
features east of the nucleus.
{\bf MC2\,1635+119} ({\it bottom middle}): Spectacular interleaved shells occur at $r$ $\sim$ 5-12\,kpc.
An arc-like feature extends out to $\sim$ 32\,kpc ({\it inset}). The shell structure is seen more prominently
when subtracting a PSF+host galaxy model as fitted by GALFIT ({\it bottom right}).
}\label{final}
\end{figure}

The ubiquity of massive black holes (BHs) in the center of galaxies shows that more 
than the mere presence of a massive BH is needed to trigger the activity observed in AGNs. 
Mergers are promising candidates and essential in at least one subclass of AGNs, ultra-luminous infrared galaxies
(\cite[Canalizo \& Stockton 2001]{can01}).
To study the relevance of mergers for the fueling of classical QSOs, we obtained
very deep (5 orbits) HST/ACS images (F606W) of five QSOs residing in host galaxies that
have been classified as undisturbed ellipticals in earlier studies (\cite[Dunlop et al. 2003]{dun03}).
For all objects, deep Keck spectroscopy revealed major starburst episodes 
($\sim$1-2 Gyr; \cite[Canalizo et al. 2006]{can06}).
For four out of the five objects,
our images reveal dramatic signs of interactions such as shells, tidal tails,
and other fine structure (Fig.~\ref{final}), suggesting that a large fraction of QSO host
galaxies may have experienced a relatively recent merger event. 
One spectacular example of shell structure is MC2 1635+119. 
In numerical simulations, the observed shells can be produced in a 
minor merger event which may also have triggered the AGN activity. 
However, we cannot exclude other scenarios such as a major merger (\cite[Canalizo et al. 2007]{can07}). 
Although all five QSO host galaxies are dominated by a de Vaucouleurs profile, 
an exponential profile contributes up to 26\% to the total host galaxy light.
The fine structure seen in four of the five QSOs
contributes up to 6\% to the total luminosity of the host galaxy (Bennert et al. 2007,
in preparation). 
However, the question remains whether the QSO host galaxies are truly
distinct from inactive ellipticals or whether we can find 
similar fine structure hinting a recent merger event.
We selected a control sample of elliptical galaxies from the HST archive.
We currently have $\sim$90 good candidates, including nine with spectroscopic redshifts. 
So far, none of the nine ellipticals has shown the spectacular fine structure found in the 
QSO hosts, although some have apparent companions.

\begin{acknowledgments}
This work was supported in part under proposals GO-10421 and AR-10941 by 
NASA through a grant from the Space Telescope Science Institute, which is 
operated by the Association of Universities for Research in Astronomy, Inc., 
under contract NAS5-26555. Additional support was provided by the National 
Science Foundation, under grant number AST 0507450. 
BJ acknowledges support by the Grant No. LC06014 of the Czech Ministry
of Education and by the Research Plan No. AV0Z10030501 of the Academy
of Sciences of the Czech Republic.
\end{acknowledgments}

\end{document}